\documentstyle[epsfig,alife8]{article}

\title{Selection of Catalysts through Cellular Reproduction}
\author{\aindx{Naoaki Ono}{Ono}$^{1}$
\and Takashi Ikegami$^{2}$\\ \mbox{}\\ 
$^{1}$ATR, Human Information Science Laboratories,  2-2-2¡¡Hikaridai,\\
"Keihanna Science City" (Seika-cho, Soraku-gun), Kyoto 619-0288, Japan\\
$^{2}$Institute of Physics, Graduate School of Arts and Sciences,\\ 
University of Tokyo, 3-8-1 Komaba Tokyo 153-8902, Japan}

\begin{document} 
\maketitle 
\begin{abstract} 
A series of simulation  studies ~\cite{Ono1999,Ono2001}
show that a proto cell 
spontaneously emerges from a chemical soup by 
acquiring membrane structures. In 2-dimensional space, the
emergence of proto cells is followed by 
the reproduction  of cells.
A major unsolved problem is the evolution of proto cells; how 
the proto cells evolve into modern cells with higher functionalities.
Here we examine, as the first step, the evolution of
catalysts within the proto-cells. Catalytic chemicals have 
different catalytic activity in generating membrane chemicals.
We show that cells with higher activity of membrane production evolve
through cellular selection.
\end{abstract}

\section{Introduction} 
It is widely accepted that the origin of
life was a set of molecules that catalyzed the reproduction
of each other.  However, when we consider
the evolution of such primitive chemical systems, 
the compartmentalization of molecules is indispensable for 
establishing the co-evolution of cooperative catalytic reactions and
protecting them from parasites that would
spoil the evolution~\cite{Szathmary1997}.
Though it is difficult to know about the structure of
primitive cells because there remain few
physical records of the earliest living cells,
there have been various theoretical approaches to understanding
the emergence and evolution of proto-cell systems.

G\'anti proposed a minimal model of primitive self-maintaining
cells named ``chemoton''~\cite{Ganti1975,Ganti1997}.  It is
composed of (1)~a metabolic system of autocatalytic molecules,
(2)~self-replicating molecules that inherit genetic information
and (3)~a self-organizing membrane molecule to enclose the whole
system.  This system maintains itself by consuming resources and 
discharging waste into the environment. 
It can be easily imagined that if the reproduction of the cell is 
appended to this system, it would be a primitive unit 
of Darwinian selection and evolve into more stable structures.

It should be stressed that a cell defines itself as an
individual by producing a membrane that distinguishes itself from
the outside.  Maturana and Varela pointed out that this is a
unique feature of living organisms, and named it
``autopoiesis''~\cite{Maturana1980}.  To demonstrate the
self-maintenance of an autopoietic structure, abstract
computational models of an autopoietic cell based on a Cellular
Automaton were proposed (originally by Varela~\cite{Varela1974},
and re-implemented by Zeleny~\cite{Zeleny1977} and by
McMullin~\cite{McMullin1997}).  
Breyer and McCaskill introduced
the metabolism of a catalyst into this model~\cite{Breyer1998}.  
It was also shown that an autopoietic proto-cell can reproduce
itself automatically~\cite{Ono1999,Ono2001}. 
Speroni di Fenizio and Dittrich proposed another approach to
represent proto-cells that are embedded in a triangular 
planar graph~\cite{Speroni2001}.

The remaining question is ``How was the {\it first} cell
organized?''  
Answering this question will give us the first step in
understanding the emergence of higher order structures in 
life's evolution.
This paper consists of two parts. The computational
algorithm of the model is explained in detail in the first part.
We introduce a Lattice Artificial
Chemistry~(LAC) model that simulates the chemical reactions and
spatial interactions of abstract chemicals.
In the second part,
an emergence of a proto-cell from a non-organized initial state, 
its reproduction and 
the selection of inside catalysts through the cell 
reproduction are reported in order.


\section{Lattice Artificial Chemistry} 
Our model is based on
discrete and stochastic dynamics, which is extended from 
a lattice-gas model.  Chemicals
are represented by particles on reaction
sites that are arranged as a two-dimensional triangular lattice.
Note that any number of particles can be placed on a single
site.  
The vector ${\mathbf{n}(\mathbf{x})} = (n_{1}({\mathbf{x}}),
n_{2}({\mathbf{x}}), \dots , n_{m}({\mathbf{x})})$ gives the number
of each type of particles on the site $\mathbf{x}$. $N_{i}$
gives the total amount of $i$-th particles in the system.

Chemical reactions are expressed by the probabilistic 
transition of particle types.  
The diffusion of chemicals is expressed by random walks of particles on
the sites.  These transition probabilities are given as the
products of the associated rate coefficients and the following
function of the potential change $\Delta E$, 
\begin{eqnarray}
  f(\Delta E) & = & \frac{\Delta E}{e^{\beta \Delta E}-1}.
\end{eqnarray} 
where $\beta$ represents 
the inverse of the product of the Boltzmann constant and temperature  
(note that, $f(\Delta E)/f(-\Delta E) = e^{-\beta \Delta E}$). In the
simulations reported hereafter, the value of $\beta$ is normalized
and fixed to 1.
\subsection{Hydrophobic Interaction} 

The probabilities of random
walks of particles are biased according to the gradient of the
potential $\Psi(\mathbf{x})$ which is given by summing up the
interaction from all particles in the same and adjacent sites.
The probability $p$ with which a particle $i$ moves from a
site $\mathbf{x}$ to $\mathbf{x'}$ is calculated as follows,
\begin{eqnarray}
  \Psi_i({\mathbf{x}}) & = & \sum_{|{\mathbf{x'}-\mathbf{x}}| \leq 1} 
    \sum_{j} \psi_{ij}({\mathbf{x'}-\mathbf{x}}) n_i({\mathbf{x}})\\
  p_i({\mathbf{x}\rightarrow
    \mathbf{x'}}) & = & {\mathit{Dif}_i} \>
    f( \Psi_i({\mathbf{x'}}) -\Psi_i({\mathbf{x}}))
\end{eqnarray} 
where $\Psi_i(\mathbf{x})$ denotes the potential of particle $i$ in the site $\mathbf{x}, $$\mathit{Dif}_i$ denotes the diffusion
coefficient of particle $i$, and $\psi_{ij}(d\mathbf{x})$ denotes
the interaction on particle $i$ from particle $j$.
Diffusion coefficients depend on the species of the particles.
Autocatalytic and membrane particles are assumed to be larger
molecules so that their diffusion coefficients are smaller than
those of other particles ($\mathit{Dif}_{A_i}=\mathit{Dif}_{M}=0.003$,
$\mathit{Dif}_{\mathit{others}}=0.01$).  

To simulate the formation of membranes, we introduce 
hydrophobic interactions between particles.  
First, the particles are 
grouped into three classes: hydrophilic, hydrophobic and neutral.  
In general, all particles repel each other,  but repulsion
between hydrophilic and hydrophobic particles is much stronger than
that between other particles so that phase separation between
different classes of particles takes place.
On the other hand, 
neutral particles do not repel other particles very strongly so that
they can diffuse more freely.

\begin{figure}[htb]
\begin{center}
  \begin{tabular}{cc}
    \epsfig{figure=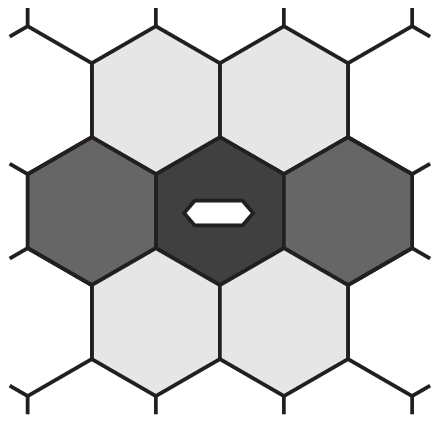,height=1.2in}&
    \epsfig{figure=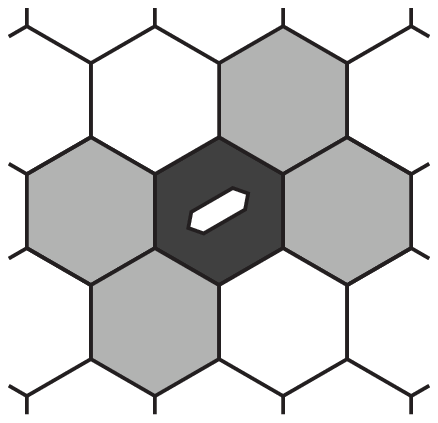,height=1.2in}\\
    (a) $\mathbf{M}^0$ & (b) $\mathbf{M}^{\pi/6}$
  \end{tabular}
  \caption{Illustration of repulsion around a particle $\mathbf{M}$.
  The honeycomb cells represent the
  lattice sites.  The depth of gray shade corresponds to the magnitude of
  repulsion against a hydrophilic particle on the site from the
  particle $\mathbf{M}$ on the center site.   
  The repulsion becomes stronger on the dark gray sites than 
  on the light gray sites.}
  \label{repulsion1} 
\end{center} 
\end{figure}

Next, we assume that hydrophobic particles $\mathbf{M}$ are 
anisotropic. Namely, the repulsion
around a particle $\mathbf{M}$ depends on its orientation and the
configuration of the particles as illustrated in 
Fig.~\ref{repulsion1}.
There are specific directions in which the repulsion becomes strong,
while the repulsion becomes weak in the other directions
Taking its symmetry
into account, a particle $\mathbf{M}$ can rotate to six
different orientations ($\mathbf{M}^0$, $\mathbf{M}^{\pm \pi/6}$,
$\mathbf{M}^{\pm \pi/3}$, $\mathbf{M}^{\pi/2}$) stochastically according
to the gradient of the potential as follows,
\begin{eqnarray} 
  p_{{\mathbf{M}}^{k \rightarrow {k'}}}
  (\mathbf{x}) & = & {\mathit{Rot}} \>
  f(\Psi_{{\mathbf{M}}^{k'}}({\mathbf{x}}) -
  \Psi_{{\mathbf{M}}^{k}}({\mathbf{x})}) 
\end{eqnarray} 
where $\mathit{Rot}=0.01$ denotes the rotation coefficient.
The repulsion between two particles $\mathbf{M}$ becomes strong
when their orientations are different, so that they tend 
to align in the same orientation. 
According to these interactions, particles $\mathbf{M}$ 
gather together to form stretched clusters. 
We call these stretched structures of particle $\mathbf{M}$ ``membrane''.
Though the characters of membranes, such as flexibility, 
depends on these values,
the formation of membranes can be observed in a wide range of parameters.
The detail values of repulsion $\psi_{ij}(d\mathbf{x})$ which
are arbitrary chosen for the following experiments 
are listed in Tables 1a and 1b.


\begin{table}[thb]
\begin{center}
  \caption{a. Repulsion between isotropic particles.}
  \label{repulsiontable1}
  \begin{tabular}{c|c|c|c}
    \multicolumn{2}{c|}{} &
    \multicolumn{2}{c}{position} \\ \cline{3-4}
    \multicolumn{2}{c|}{particles} & $dr=0$ & $dr=1$ \\ \hline
    hydrophilic & hydrophilic & 0.0100 & 0.0033 \\
    \cline{2-4} {} & neutral & 0.0010 & 0.0003 \\ \hline
    neutral & neutral & 0.0010 & 0.0003 \\
  \end{tabular}
\end{center} 
\end{table}

\subsection{Chemical Reaction} 
We introduce a simple metabolic system of autocatalytic
particles. Consider that there are various species of self-replicating
particles, 
and some of them have the ability to catalyze the production of membrane
particles. 
Resources of these particles are supplied from some
external source homogeneously.

\begin{figure}[tbh]
\begin{center}
  \begin{tabular}{c}
    \epsfig{figure=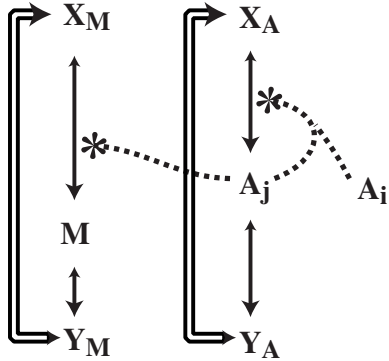,width=2in}\\
  \end{tabular}
  \caption{Schematic drawings of reaction paths.
  An autocatalyst ($\mathbf{A_i}$) catalyzes
  the reproduction of another particle $\mathbf{A_j}$ from
  a resource particle ($\mathbf{X_A}$) that has a higher chemical 
  energy. It also catalyzes the production of a membrane particle
 ($\mathbf{M}$) from another resource particle ($\mathbf{X_M}$). 
  All particles decay
  into waste particles ($\mathbf{Y_A}$ and $\mathbf{Y_M}$, respectively),
  spontaneously, however,
  an external energy supply recycles $\mathbf{Y_A}$ and
  $\mathbf{Y_M}$
  into $\mathbf{X_A}$ and $\mathbf{X_M}$, respectively.
  The number of total particles is preserved.}
  \label{paths1} 
\end{center} 
\end{figure}

Figure~\ref{paths1} illustrates the reaction paths.
The probabilities of chemical reactions depend on the
enthalpy change along with the transition, as follows,
\begin{eqnarray}
  \Delta H_{ij} &=& \Delta H_j - \Delta H_i \\
  p_{i \rightarrow j}(\mathbf{x}) & = & k_{ij}({\mathbf{x}})
    \> f(\Delta H_{ij} + \Psi_{j}({\mathbf{x}}) -
    \Psi_{i}(\mathbf{x})) 
\end{eqnarray} 
where $\Delta H_{ij}$ denotes the enthalpy change that
is given by the difference in the formation enthalpy listed in
Table~\ref{enthalpytable1}.\\
\twocolumn[
\begin{center}
Table 1.b Repulsion between hydrophobic and other particles
  \begin{tabular}{c|c|c|c|c|c}
\multicolumn{2}{c|}{} &
\multicolumn{4}{c}{position} \\ \cline{3-6}
\multicolumn{2}{c|}{particles} & $dr=0$ &
\multicolumn{3}{c}{$dr=1$} \\ \cline{4-6}
\multicolumn{2}{c|}{} & {} & $\theta=0,\pi$ &
$\theta=\pi/3,-2\pi/3$ & $\theta=2\pi/3,-\pi/3$ \\
\hline
{} & hydrophilic & 0.2000 & 0.1600 & 0.0200 & 0.0200 \\
\cline{2-6}
{} & neutral & 0.0010 & 0.0008 & 0.0001 & 0.0001 \\
\cline{2-6}
$\mathbf{M}^{0}$ &$\mathbf{M}^{0}$ & 0.0100 & 0.0033 &
0.0033 & 0.0033 \\
\cline{2-6}
{} & $\mathbf{M}^{\pi/6}$ & 0.0777 & 0.0259 & 0.0259 &
0.0259 \\
\cline{2-6}
{} & $\mathbf{M}^{\pi/3}$ & 0.1433 & 0.0477 & 0.0477 &
0.0477 \\
\cline{2-6}
    {} & $\mathbf{M}^{\pi/2}$ & 0.2100 & 0.0700 & 0.0700 &
      0.0700 \\ \hline
    {} & hydrophilic & 0.2000 & 0.1000 & 0.1000 & 0.0000 \\
      \cline{2-6}
    $\mathbf{M}^{\pi/6}$ & neutral & 0.0010 & 0.0005 &
      0.0005 & 0.0000 \\ \cline{2-6}
    {} & $\mathbf{M}^{\pi/6}$ & 0.0100 & 0.0033 & 0.0033 & 
      0.0033 \\ \cline{2-6}
    {} & $\mathbf{M}^{-\pi/3}$ & 0.2100 & 0.0700 & 0.0700 &
      0.0700 \\
  \end{tabular}
\end{center}
\vspace{0.2cm}
]
$k_{ij}(\mathbf{x})$ denotes the coefficient of reaction 
$i \leftrightarrow j$ that may depend on the number of catalysts on
the site.  Note that the effects of the
interactive potential, namely, the effects of
hydrophilic/hydrophobic environments are also taken into account
here, therefore, for example, 
it becomes more difficult to synthesize
a hydrophilic particle in a hydrophobic environment.

\begin{table}[htb] 
\begin{center}
  \caption{Formation enthalpy}
  \label{enthalpytable1}
  \begin{tabular}{c|c|c|c|c}
    particle & {$\mathbf{X_A}$, $\mathbf{X_M}$} & 
    {$\mathbf{A_i}$} & {$\mathbf{M}$} &
    {$\mathbf{Y_A}$, $\mathbf{Y_M}$} \\ \hline
    $\Delta H_i$ & 12.0 & 6.0 & 4.0 & 0.0
  \end{tabular} 
\end{center} 
\end{table}

There are ten species of autocatalytic particles ($\mathbf{A_0}$
$\dots$ $\mathbf{A_9}$). 
An autocatalytic particle $\mathbf{A_i}$ catalyzes the
replication of another particle $\mathbf{A_j}$ using it as a template and
consuming a resource particle ($\mathbf{X_A}$).
\begin{eqnarray}
  \mathbf{A_i} + \mathbf{A_j} &\longleftrightarrow&
  \mathbf{A_i}\mathbf{A_j} \\
  \mathbf{A_i}\mathbf{A_j} + \mathbf{X_A} &\longleftrightarrow&
  \mathbf{A_i}\mathbf{A_j} + \mathbf{A_j} 
\end{eqnarray}
There is a probability of mutation $\mu$ with which a particle 
$\mathbf{A_i}$ mutates to $\mathbf{A_{i\pm 1}}$ when it is reproduced.
Assuming that the rate of the first reaction is much faster than
that of the second one, the rate coefficients between $\mathbf{X_A}$ and
$\mathbf{A_j}$ can be given as follows,
\begin{equation}
  n'_{A_i}({\mathbf{x}}) = \mu n_{A_{j-1}}({\mathbf{x}})+
    (1-2 \mu) n_{A_j}({\mathbf{x}})+ \mu n_{A_{j+1}}({\mathbf{x}})
\end{equation} 
\begin{eqnarray}
  k_{X_A \leftrightarrow A_j}({\mathbf{x}}) &=& k_A + C_A 
    n'_{A_i}({\mathbf{x}}) \sum_i n_{A_i}({\mathbf{x}}) 
\end{eqnarray} 
where $n_{A_i}({\mathbf{x}})$ denotes the number of particles
$\mathbf{A_i}$ on the site $\mathbf{x}$, and
$k_A$ denotes the rate of spontaneous reaction. 
Note that all autocatalysts share a common
catalytic activity $C_A$ and catalyze the replication of
each other equally.

An autocatalytic particle also catalyzes 
the production of a membrane 
particle ($\mathbf{M}$) from a resource ($\mathbf{X_M}$).  
The activity ($C_{M_i}$) depends on the species.
The catalytic activity of each species $\mathbf{A_i}$ is
given by the following equation, namely, the activity
of particle $\mathbf{A_i}$ is $i$-th times larger 
than that of particle $\mathbf{A_1}$, so that the rate 
coefficients between $\mathbf{X_M}$ and $\mathbf{M}$ are 
given as follows,
\begin{eqnarray}
  C_{M_i} & = & C_M \times i \\
  k_{X_M \leftrightarrow M}(\mathbf{x}) &=& k_M +
    \sum_i C_{M_i} n_{A_i}(\mathbf{x}),
\end{eqnarray} 
where $C_M$ is a given constant, 
and $k_M$ denotes the rate of spontaneous reaction.

These particles naturally decay into waste particles
($\mathbf{Y_A}$ and $\mathbf{Y_M}$, respectively) at a constant
rate $k_Y$.  However, we introduce an external
source that supplies resources.  To preserve the total number of
particles, the resource supply is expressed by the exchange from
waste to resource particles.
Thus the transition coefficients are given as follows,
\begin{eqnarray}
  k_{A_j \leftrightarrow Y_A} = k_{M \leftrightarrow Y_M}
    &\equiv& k_Y \\
  k_{X_A \rightarrow Y_A} = k_{X_M \rightarrow Y_M} 
    &\equiv& k_Y \\
  k_{Y_A \rightarrow X_A} = k_{Y_M \rightarrow X_M} 
    &\equiv& k_Y + S_X.  
\end{eqnarray}
Due to the term $S_X$, the whole system is kept in a
non-equilibrium state.  
The last particle ($\mathbf{W}$) represents water 
that does not change into other particles.
We assume that water and autocatalytic particles are 
hydrophilic particles that are repelled by membranes, and
resource and waste particles are neutral particles which
can diffuse through membranes.

The rate
coefficients of spontaneous reactions are $ k_A = k_M = 1.0
\times 10^{-8} $, $k_Y = 1.0 \times 10^{-4} $. The coefficients
of catalytic activity are $C_A = 1.0 \times 10^{-5}$ and $C_M =
1.0 \times 10^{-5}$. The mutation rate is $\mu=1.0 \times
10^{-12}$. The rate of resource supply is given a constant 
$S_X=16$.

The simulation is based on a Metropolice method.  
At each iteration, the following steps are repeated,
\begin{enumerate} 
  \item Calculate the potential of each particle.
  \item Calculate the probabilities of diffusion, rotation and
    chemical transition according to the potential difference.
  \item Change the state of particles according to the 
    probabilities synchronously.
\end{enumerate}

In the initial state, the particles are placed
randomly.  There are 30 particles on a site on average,
and the mean numbers of particles on a site are listed in
Table~\ref{initialpopulations1}.  There is a sufficient number
of resource particles and supplies to sustain metabolism.
The average production rate of membranes is set very low at first.
Catalysts with higher activity only emerge through
random mutations. 
The reaction sites are arranged as a $64
\times 64$ triangular lattice whose boundaries are periodic.

\begin{table}[htb] 
\begin{center}
  \caption{Mean numbers on a site in the initial state}
  \label{initialpopulations1}
  \begin{tabular}{c|c|c|c|c|c|c}
    particle & $\mathbf{A_0}$ & $\mathbf{A_1}$ &
    $\mathbf{A_2}$ & $\mathbf{A_3}$ &
    \multicolumn{2}{c}{$\mathbf{A_4} \dots \mathbf{A_9}$} \\ \hline
    $\overline{n}_{i}$ & 1.6 & 1.2 & 0.8 & 0.4 &
    \multicolumn{2}{c}{0.0} \\ \hline \hline
    particle & $\mathbf{X_A}$ & $\mathbf{Y_A}$ & $\mathbf{M}$ &
    $\mathbf{X_M}$ & $\mathbf{Y_M}$ & $\mathbf{W}$ \\ \hline
    $\overline{n}_{i}$ & 2.0 & 2.0 & 0.0 & 5.0 & 5.0 & 10.0
  \end{tabular} 
\end{center} 
\end{table}

\section{Simulation Results} 
The evolution of this system is roughly divided into three
characteristic stages: (1)~Chemical evolution, (2)~Emergence
of proto-cells, and (3)~Cellular evolution.

\subsection{Chemical Evolution} 
\begin{figure}[htb]
  \begin{center}
  \begin{tabular}{cc}
    \epsfig{figure=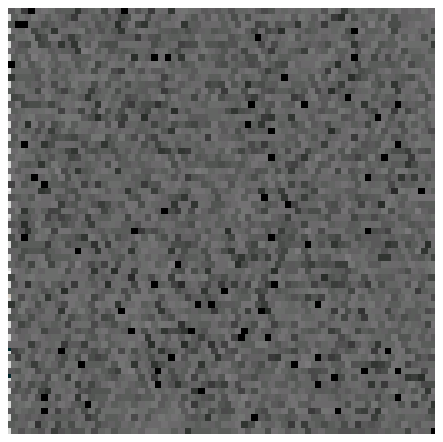,width=1.4in}&
    \epsfig{figure=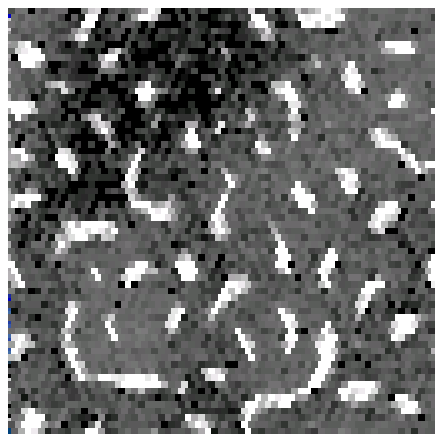,width=1.4in}\\
    (1) Initial state & (2) After 30,000 iterations \\
  \end{tabular}
  \caption{Chemical evolution.
  The white regions are dominated by particle $\mathbf{M}$.
  The depth of gray shade represents the total population of
  the autocatalysts ($\sum \mathbf{A_i}$). The black regions
  are dominated by particle $\mathbf{W}$.
  Resource and waste particles are not displayed in the
  figures. Pieces of membranes are produced by the catalysts
  which emerged through mutations.}
  \label{evolution1}
\end{center} 
\end{figure}

Fig.~\ref{evolution1}(1) shows the initial configuration.
Before cellular selection starts, the chemical evolution 
simply depends on the reproduction rate of each species. 
In this model, because
they share the same reproduction rate, the evolution 
is mostly driven by mutations and random fluctuation.
At first, the largest part of the autocatalytic particles is
$\mathbf{A_0}$ which does not produce membrane particles.
Table~\ref{populations2} shows a profile of the population after
30,000 iterations for a single run.
However, as the populations of other species increase,
small pieces of membranes are gradually formed.
(Fig.~\ref{evolution1}(2)).

\begin{table}[htb] 
\begin{center}
  \caption{A profile of the population of particles after 30,000
  iterations.}
  \label{populations2}
  \begin{tabular}{c|c|c|c|c|c}
    particle & $\mathbf{A_0}$ & $\mathbf{A_1}$ &
    $\mathbf{A_2}$ & $\mathbf{A_3}$ & $\mathbf{A_4}$ \\ \hline
    $\overline{n}_{i}$ & 1.00 & 0.60 & 0.42 & 0.20 & 0.05 \\ \hline \hline
    particle & $\mathbf{A_5}$ & $\mathbf{A_6}$ & $\mathbf{A_7}$ 
      &$\mathbf{A_8}$ & $\mathbf{A_9}$ \\ \hline
   $\overline{n}_{i}$ & 0.07 & 0.08 & 0.05 & 0.04 & 0.06 \\ \hline \hline
    particle & $\mathbf{X_A}$ & $\mathbf{Y_A}$ & 
      $\mathbf{M}$ & $ \mathbf{X_M}$ & $\mathbf{Y_M}$  \\ \hline
    $\overline{n}_{i}$ & 3.00 & 4.43 & 1.57 & 3.68 & 4.77 
  \end{tabular} 
\end{center} 
\end{table}

\subsection{Emergence of Proto-cells} 
Once membranes are formed,
they begin to restrict the diffusion of catalysts. Thus, membranes can
keep the local population and also their reaction rate high.   
As resource
particles are consumed faster in such regions, resource particles
diffuse into these regions according to the gradient of the 
population. It increases their reaction rate more. 
Due to this osmotic competition for resources,
a small difference in the population of
autocatalysts between the two sides of the membrane becomes larger.

\begin{figure}[htb] 
\begin{center}
  \begin{tabular}{cc}
    \epsfig{figure=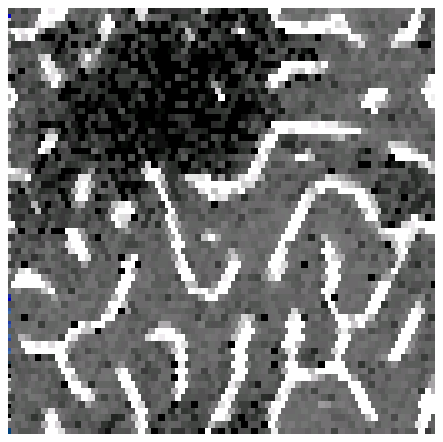,width=1.4in}&
    \epsfig{figure=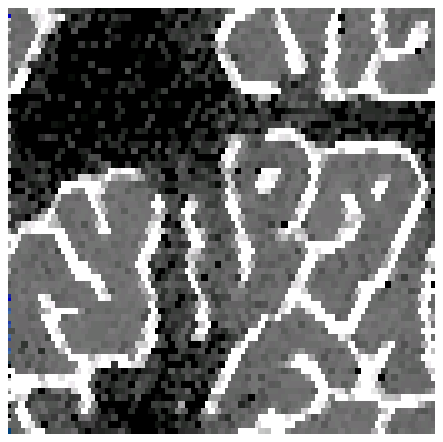,width=1.4in}\\
    (1) 60,000 iterations & (2) 120,000 iterations \\
    \epsfig{figure=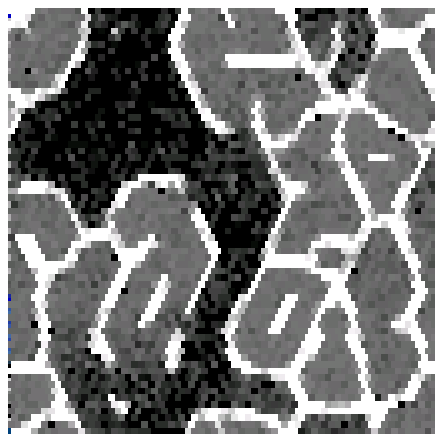,width=1.4in}&
    \epsfig{figure=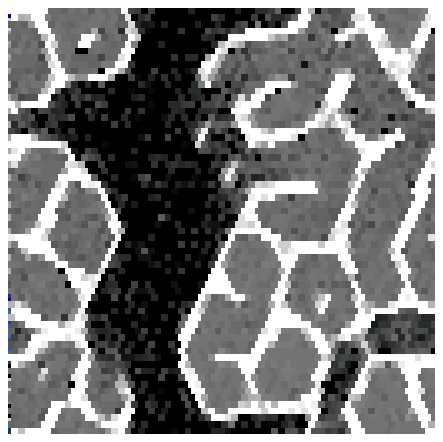,width=1.4in}\\
    (3)  180,000 iterations & (4)  240,000 iterations \\
  \end{tabular}
  \caption{Emergence of Proto-cell structures. As the membranes
  grow,	competition for resources between regions separated by
  membranes takes place. The regions differentiate into two states.
  In some regions that are enclosed by membranes, the density of 
  autocatalysts stays high.
  In the other regions, their density becomes almost zero.}
  \label{evolution2}
\end{center} 
\end{figure}

When the density of resources becomes too 
low in some regions, autocatalysts are 
no longer able to sustain their replication.  Autocatalysts and membrane 
particles in these regions decay. 
At last, most regions become inactive, namely, filled with only resource 
and waste particles (and water), while there remain some 
active regions in which autocatalysts keep reproducing.
These structures maintain themselves autopoietically,
namely, the autocatalysts inside them reproduce themselves and 
metabolize the membrane particles to maintain their membranes.
We call this structure a ``proto-cell'' hereafter.

\subsection{Cellular Reproduction} 
\begin{figure}[htb] 
\begin{center}
  \begin{tabular}{cc}
    \epsfig{figure=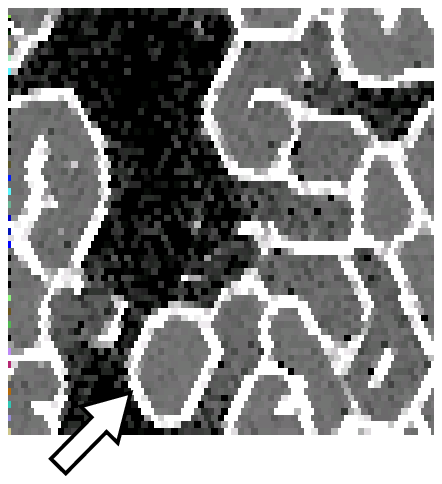,width=1.4in} &
    \epsfig{figure=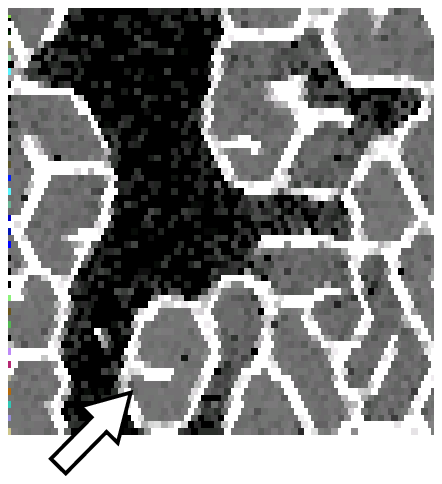,width=1.4in} \\
    (1)  300,000 iterations & (2)  309,000 iterations \\
    \epsfig{figure=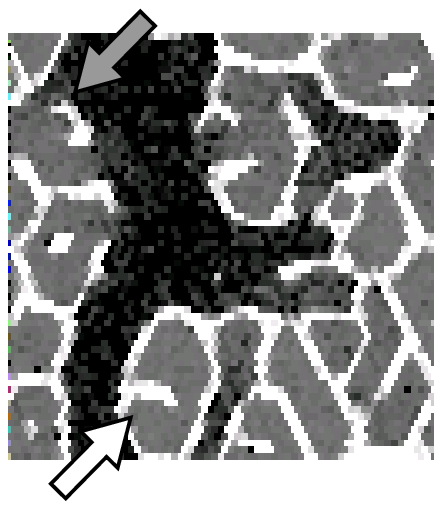,width=1.4in} &
    \epsfig{figure=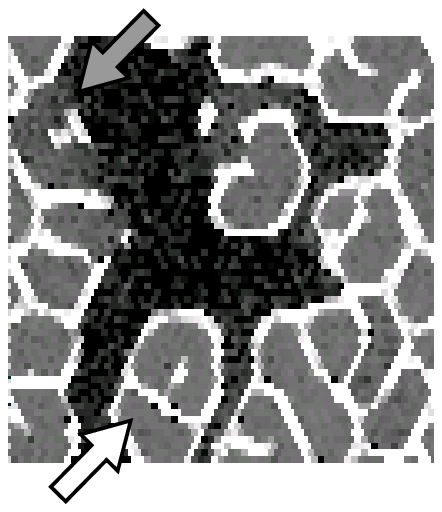,width=1.4in} \\
    (3)  318,000 iterations & (4)  327,000 iterations \\
  \end{tabular}
  \caption{Reproduction of proto-cell. Snapshots from 300,000 
  iterations to 327,000 iterations. The proto-cell indicated by the white 
  arrow grows gradually in size. 
  When it becomes too large, another membrane appears inside it.
  At last, it divides the cell into daughter cells.
  On the other hand, the cell indicated by the gray arrows could not 
  produce enough membrane particles to keep its membrane.
  Catalysts in the cell diffuse away through the defect of the membrane.
  }
  \label{evolution3}
\end{center}
\end{figure}

A proto-cell can not only maintain itself but can reproduce itself.
Figure~\ref{evolution3} shows
snapshots of the reproduction process.  
A proto-cell structure grows in size by assimilating resource
particles from neighboring regions.  As it grows, it comes to
produce more membrane particles than it needs to maintain its
membrane.  When it reaches a certain size, surplus membrane
particles begin to form another membrane inside the cell. This
divides the mother cell into a few daughter cells.  The daughter
cells can continue to grow and reproduce recursively.

On the other hand, if a cell fails to maintain its membrane, the
density of catalysts lowers quickly due to diffusion through the
defect of membrane.
When the density becomes too low, the catalysts
can not sustain the metabolism any
longer, and the whole structure finally disappears.
This mutual dependency between membrane and metabolic system is
essential to the evolution of proto-cells.

\subsection{Selection of Catalysts Through Cellular Reproduction}
\begin{figure}[htb] 
\begin{center}
  \begin{tabular}{c}
    \epsfig{figure=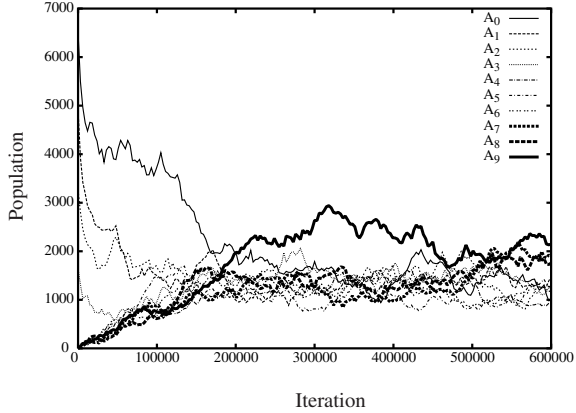,width=3in}
  \end{tabular}
  \caption{Evolution of the population of autocatalysts.
  At first, the population of 
  $\mathbf{A_0}$, which does not produce membrane
  particles, is the largest. However, other species 
  arise through mutation, and after the formation of proto-cells, 
  species which can produce membranes have
  an advantage over $\mathbf{A_0}$ (see the thick lines which denote 
  species $\mathbf{A_7}$ -- $\mathbf{A_9}$).}
  \label{timeseries1}
\end{center} 
\end{figure}

Figure~\ref{timeseries1} shows the evolution of the population
of catalysts. 
If there is no spatial structure,
the populations of catalysts just randomly drift around 
the equilibrium where all populations are equal.
However, once the differentiation into active and inactive regions
has advanced, it becomes difficult to sustain metabolism without 
membranes, because the density of autocatalysts lowers fast.

Note that when a proto-cell divides itself, the composition of
the population of replicators contained in the cell is roughly
inherited by the daughter cells. 
This implies that catalysts that can sustain a proto-cell more stably
are selected regarding a proto-cell as a new unit 
of Darwinian evolution.
Therefore, due to cellular selection, the populations of catalysts are
biased toward  higher membrane production
activity against the random drift.
The populations of catalysts and their relative amount 
($\rho_{i} = \overline{n_i} / \sum \overline{n_j} \times 100$)
after 600,000 iterations for the same run as the previous one 
are listed in Table~\ref{populations3}. 


\begin{table}[htb] 
\begin{center}
  \caption{A profile of the populations of catalysts after
  600,000 iterations.}
  \label{populations3}
  \begin{tabular}{c|c|c|c|c|c}
    particle & $\mathbf{A_0}$ & $\mathbf{A_1}$ &
      $\mathbf{A_2}$ & $\mathbf{A_3}$ & $\mathbf{A_4}$\\ \hline
    $\overline{n}_{i}$ & 0.24 & 0.36 & 0.28 & 0.30 & 0.23 \\ \hline 
      $\rho_{i}$(\%) &  7.0 & 10.5 & 8.1 & 8.7 & 6.7 \\ \hline \hline
    particle & $\mathbf{A_5}$ & $\mathbf{A_6}$ &
      $\mathbf{A_7}$ & $\mathbf{A_8}$ & $\mathbf{A_9}$\\ \hline
    $\overline{n}_{i}$ & 0.31 & 0.31 & 0.46 & 0.43 & 0.52 \\ \hline 
      $\rho_{i}$(\%) &  9.0 & 9.0 & 13.4 & 12.5 & 15.1
  \end{tabular} 
\end{center} 
\end{table}

To make the effects of the cellular selection more clear, we investigated
evolution under a lower resource supply.  
The initial configuration and the parameters are the same as those of
the previous run.
After the formation of the proto-cell structures (after 300,000
iterations),  
the rate of resource supply $S_X$ was decreased to
$S_X \times 0.75$.
Because the density of resource particles decreases, catalysts have to
keep their density high to maintain a sufficient reproduction rate more
tightly, and a cell that fails to maintain its membrane disappears faster.
The pressure to acquire higher membrane production activity
becomes stronger.
Table~\ref{populations4} shows the populations after 900,000
iterations for another run.
The dominance of catalysts with higher activity is clearer.

\begin{table}[htb] 
\begin{center}
  \caption{A profile of populations after 900,000 iterations under
  the lower resource supply.}
  \label{populations4}
  \begin{tabular}{c|c|c|c|c|c}
    particle & $\mathbf{A_0}$ & $\mathbf{A_1}$ &
      $\mathbf{A_2}$ & $\mathbf{A_3}$ & $\mathbf{A_4}$\\ \hline
    $\overline{n}_{i}$ & 0.19 & 0.10 & 0.11 & 0.16 & 0.13 \\ \hline 
      $\rho_{i}$(\%) &  8.6 & 4.5 & 5.0 & 7.3 & 5.9 \\ \hline \hline
    particle & $\mathbf{A_5}$ & $\mathbf{A_6}$ &
      $\mathbf{A_7}$ & $\mathbf{A_8}$ & $\mathbf{A_9}$\\ \hline
    $\overline{n}_{i}$ & 0.18 & 0.25 & 0.20 & 0.42 & 0.46 \\ \hline 
      $\rho_{i}$(\%) &  8.2 & 11.4 & 9.1 & 19.1 & 20.9
  \end{tabular} 
\end{center} 
\end{table}

\begin{figure}[bth] 
\begin{center}
 \begin{tabular}{c}
  \epsfig{figure=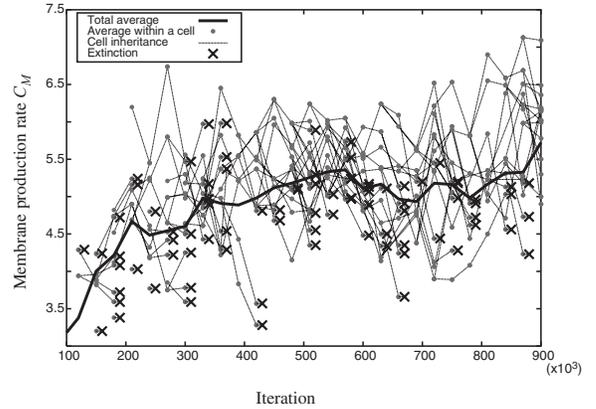,width=3in}
 \end{tabular} 
\end{center} 
 \caption{Evolution of the mean activity of
 membrane production ($\overline{C_{M_i}}$). A cell with lower
  membrane production breaks more often because its membrane
  is weaker.}
 \label{evolution4}
\end{figure}

To see the detailed process of cellular selection, 
the evolution of the mean activity 
($\overline{C_{M_i}} = \sum C_{M_i} N_{A_i} / \sum N_{A_i}$)
of membrane production is shown in Fig.~\ref{evolution4}.
Within a cell, because competition among the autocatalysts is neutral,
the mean activity of each cell fluctuates randomly.
But among the cellular assembly, 
a proto-cell in which membrane
production activity is low becomes extinct more often.
As a result, cells with higher catalytic activity outperform the
lower ones and the total average of $\overline{C_{M_i}}$
gradually increases.

\begin{figure}[htb] 
\begin{center}
  \begin{tabular}{cc}
    \epsfig{figure=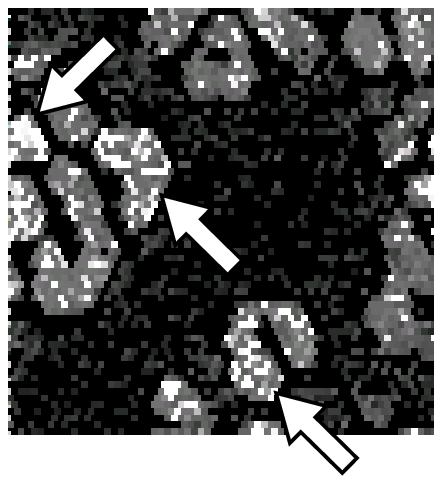,width=1.4in} &
    \epsfig{figure=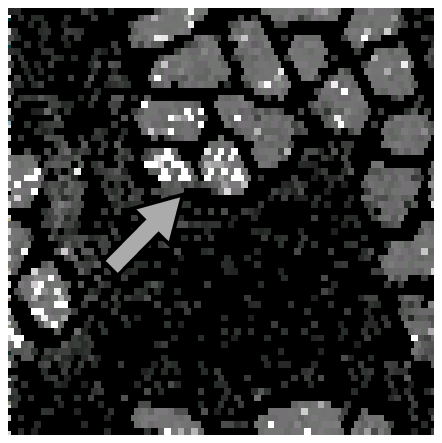,width=1.4in} \\
     (1)  780,000 iterations & (2)  900,000 iterations
  \end{tabular}
  \caption{Invasion of parasitic catalysts. The white sites 
  are where the mean catalytic activity is low.}
  \label{parasites1} 
\end{center} 
\end{figure}

Though there often appear cells with very high activity of
membrane production, the $\overline{C_{M_i}}$ of these cells 
soon drops again. 
These drops are caused by the invasion of ``parasitic'' catalysts.
Namely, the evolved cells are dominated by catalysts that have higher
activity and produce enough membrane particles, however,
the catalysts with lower activity, e.g., $\mathbf{A_0}$, can always
emerge through mutation and increase through the random 
fluctuation within these cells.
Snapshots of the process of cellular selection are
shown in Fig.~\ref{parasites1}.
At 780,000 iterations,
the cells indicated by the white arrows are deeply infected by parasites. 
It becomes difficult for these cells to maintain their membranes
and they disappear before 900,000 iterations.
The cells at 900,000 iterations are produced from survived cells
that have higher catalytic activity, 
but, there are cells that are newly invaded by parasites (indicated by
the gray arrow).
This result indicates that these proto-cells have limited ``life spans''.
A proto-cell has to keep dividing itself to escape from parasites, 
otherwise, the parasitic catalysts increase in it before long.


\begin{figure}[htb] 
\begin{center}
  \begin{tabular}{c}
    \epsfig{figure=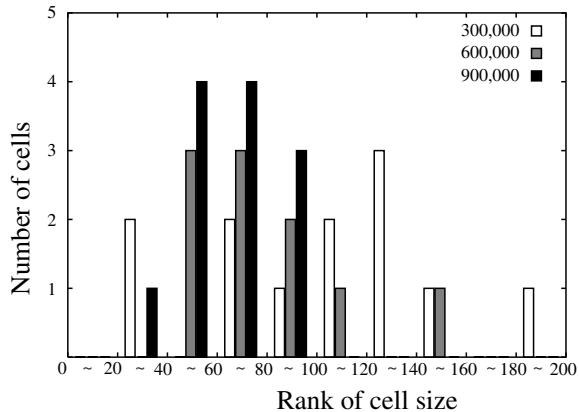,width=3in}
  \end{tabular}
  \caption{The distribution of cell sizes after 300,000, 600,000 and 
  900,000 iterations.
  The standard deviations of the sizes are 
  47.52, 26.58,and  24.26, respectively.}
  \label{sizedistribution1} 
\end{center} 
\end{figure}

As the cellular evolution proceeds, it is observed that the size of
the cells becomes smaller. 
Figure~\ref{sizedistribution1} shows a histogram of cell sizes
(namely, the number of sites within the membranes) for the same run,
after 300,000, 600,000 and 900,000 iterations.
The mean sizes of cells at each iteration are 103.8, 82.7, and 66.2,
respectively.
When the mean activity of the membrane production 
of a cell increases, it can maintain its membrane with fewer catalysts,
and it can divide itself faster. 
However, it is sometimes observed that a cell whose size is too small
fails to maintain itself.
When a cell produces too many membrane particles or the size of 
cell become too small, the rate of replication of the autocatalysts
decreases, because 
the hydrophobic environment suppress the synthesis of hydrophilic 
particles.
A certain optimum size of the cell 
will be achieved through the selection of catalysts.
This expectation is supported by the result that, in the 
evolved state, the dispersion of cell sizes becomes smaller.

\section{Conclusion and Discussion} 
In this article, we have
presented a model for the evolution from molecular to cellular
reproduction.  Starting from a homogeneous, random initial state,
we demonstrated the emergence of proto-cells.  This goes through
three stages: (1)~Metabolic cycles that produce membrane
molecules can arise through pre-cellular chemical evolution.
(2)~Proto-cell structures, i.e., self-maintaining structures
that maintain their own membranes by themselves emerge
spontaneously.  (3)~A proto-cell structure reproduces
itself. Because the molecules contained inside a proto-cell can pass
genetic information into daughter cells, it can be
regarded as a unit of Darwinian evolution. Cells that can
maintain themselves more stably evolve through cellular
selection.

Excess production of membrane particles generates 
a hydrophobic environment, which is less optimal for the
self-reproduction of the particles. Nevertheless, 
particles with higher catalytic ability replicate more 
via cellular level selection. As a result, the average value
of catalyzation with membrane formation becomes higher than that 
without membrane formation.
We insist here that the emergence of cellular structures 
produces a new rugged evolutionary landscape on the particle level.

The entire behavior is insensitive to parameter values 
whenever we have the conditions of,
 1) a membrane formation and 2) transportation of resource
particles through membranes. 
This robust behavior gives an advantage to our model over other
 cellular formation systems such as Grey-Scott, whose
spatio-temporal pattern is sensitive to reaction and diffusion
rates.
In our model, the size of cells and replication rates change
but their qualitative behaviors never change.
Instead, our model behavior is sensitive to the form of repulsion
potential of the membrane particles. 
In this study, we chose a set of values which provide flexible
membranes to make the organization of proto-cells easier.
We take the advantage of this sensitivity to the potential form
in order to study the evolutionary dynamics of these
characteristics of membranes of proto-cells.

In the simulations reported here, the resources supply was
homogeneous.  And the pressure to evolve membrane production
became clear when the resource supply rate was lowered.
These results suggest that, if the supply of resource is not
homogeneous, namely, if there is a small region where the resource
supply is plenty enough to sustain metabolism, but the supply is
rather poor in the other regions, pre-cellular evolution will
take place where the resource supply is high enough, and once
the proto-cell structures are acquired, they can migrate where
the resource supply is lower, which will promote the further
evolution of cellular structures.

Though the model introduced in this article is simple and
abstract, we are now extending our model to implement more
complex metabolic reactions that can produce more diverse
membrane particles and various cell types.  Our next objective
is to investigate the co-evolution of metabolic systems and cellular
structures.  Because in our model, the birth and death of cells
are all actualized through the elemental interactions of
particles, without any ad-hoc rules, the way in which the
proto-cell is organized itself can evolve.  In this sense, it
can be regarded as a basic model for the organization of a dynamical
hierarchy.  We expect that higher order structures (e.g., cell
differentiation and cooperative interaction between cells) are
yielded.

Furthermore, the dynamics of our model is based on a local
equilibrium system.  A quantitative analysis of this model may bring
insight to the evolution of primitive cells as a
non-equilibrium system from a thermo-dynamical aspect.  Along
these lines, computational models of artificial chemistry will
provide useful tools for understanding the earliest evolution of
life.

\section*{Acknowledgments} 
This work was partially supported by the COE project
(``complex systems theory of life''),
a Grant-in-aid (No. 11837003) from the Ministry of Education,
Science, Sports and Culture and afiis project (Academic
Frontier, Intelligent Information Science 2000 - 2004 Doshisha University).

\bibliographystyle{aaai} 
\bibliography{nono}

\begin{thebibliography}{}

\bibitem[\protect\citeauthoryear{Breyer, Ackermann, \&
  McCaskill}{1998}]{Breyer1998}
Breyer, J.; Ackermann, J.; and McCaskill, J.
\newblock 1998.
\newblock Evolving reaction-diffusion ecosystems with self-assembling
  structures in thin films.
\newblock {\em Artificial Life} 4:25--40.

\bibitem[\protect\citeauthoryear{G\'anti}{1975}]{Ganti1975}
G\'anti, T.
\newblock 1975.
\newblock Organization of chemical reactions into dividing and metabolizing
  units: The chemotons.
\newblock {\em BioSystems} 7:15--21.

\bibitem[\protect\citeauthoryear{G\'anti}{1997}]{Ganti1997}
G\'anti, T.
\newblock 1997.
\newblock Biogenesis itself.
\newblock {\em J. theol. Biol.} 187:583--593.

\bibitem[\protect\citeauthoryear{Maturana \& Varela}{1980}]{Maturana1980}
Maturana, H.~R., and Varela, F.~J.
\newblock 1980.
\newblock {\em Autopoiesis and Cognition: the Realization of the Living}.
\newblock D. Reidel Publishing.

\bibitem[\protect\citeauthoryear{McMullin \& Varela}{1997}]{McMullin1997}
McMullin, B., and Varela, F.~J.
\newblock 1997.
\newblock Rediscovering computational autopoiesis.
\newblock In Husbands, P., and Harvey, I., eds., {\em 4th European Conference
  on Artificial Life},  38--47.
\newblock Brighton, UK: MIT press.

\bibitem[\protect\citeauthoryear{Ono \& Ikegami}{1999}]{Ono1999}
Ono, N., and Ikegami, T.
\newblock 1999.
\newblock Model of self-replicating cell capable of self-maintenance.
\newblock In Floreano, D.; Nicoud, J.~D.; and Mondada, F., eds., {\em
  Proceedings of the 5th European Conference on Artificial Life (ECAL'99)},
  399--406.
\newblock Lausanne, Switzerland: Springer.

\bibitem[\protect\citeauthoryear{Ono \& Ikegami}{2001}]{Ono2001}
Ono, N., and Ikegami, T.
\newblock 2001.
\newblock Artificial chemistry: Computational studies on the emergence of
  self-reproducing units.
\newblock In Kelemen, J., and Sosik, S., eds., {\em Proceedings of the 6th
  European Conference on Artificial Life (ECAL'01)},  186--195.
\newblock Prague, Czech Republic: Springer.
\bibitem[\protect\citeauthoryear{Speroni di Fenizio, Dittrich, \&
  Banzhaf}{2001}]{Speroni2001}

Speroni~di~Fenizio, P.; Dittrich, P.; and Banzhaf, W.
\newblock 2001.
\newblock Spontaneous formation of proto-cells in a universal artificial
  chemistry of a planar graph.
\newblock In Kelemen, J., and Sosik, S., eds., {\em Proceedings of the 6th
  European Conference on Artificial Life (ECAL'01)},  206--215.
\newblock Prague, Czech Republic: Springer.

\bibitem[\protect\citeauthoryear{Szathm{\'a}ry \& {Maynard
  Smith}}{1997}]{Szathmary1997}
Szathm{\'a}ry, E., and {Maynard Smith}, J.
\newblock 1997.
\newblock From replicators to reproducers: the first major transitions leading
  to life.
\newblock {\em J. theor. Biol} 187:555--571.

\bibitem[\protect\citeauthoryear{Varela, Maturana, \& Uribe}{1974}]{Varela1974}
Varela, F.~J.; Maturana, H.~R.; and Uribe, R.
\newblock 1974.
\newblock Autopoiesis: The organization of living systems, its characterization
  and a model.
\newblock {\em BioSystems} 5:187--196.

\bibitem[\protect\citeauthoryear{Zeleny}{1977}]{Zeleny1977}
Zeleny, M.
\newblock 1977.
\newblock Self-organization of living systems: A formal model of autopoiesis.
\newblock {\em International Journal of General Science} 4:13--28.

\end{thebibliography}

\end{document}